\documentclass[11pt,twoside]{article}

\usepackage{asp2014}

\aspSuppressVolSlug
\resetcounters

\begin{document}

\title{Improved Acceleration of the GPU Fourier Domain Acceleration Search Algorithm}

\author{Karel~Ad\'{a}mek,$^1$ Sofia Dimoudi,$^2$ Mike Giles,$^3$ and Wesley~Armour$^1$
\affil{$^1$Oxford e-Research Centre, Department of Engineering Science, University of Oxford, Oxford, OX1 3QG, UK; }
\affil{$^2$Centre for Clinical Magnetic Resonance Research, University of Oxford, Oxford, OX3 9DU, UK;}
\affil{$^3$Mathematical Institute, University of Oxford, Oxford, OX2 6GG, UK;}
\email{karel.adamek@oerc.ox.ac.uk}
}

\paperauthor{Karel~Ad\'{a}mek}{karel.adamek@oerc.ox.ac.uk}{0000-0003-2797-0595}{University of Oxford}{Oxford e-Research Centre}{Oxford}{}{OX1 3QG}{UK}
\paperauthor{Sofia~Dimoudi}{sofia.dimoudi@cardiov.ox.ac.uk}{}{University of Oxford}{Centre for Clinical Magnetic Resonance Research}{John Radcliffe Hospital, Headington}{Oxford}{OX3 9DU}{UK}
\paperauthor{Mike~Giles}{mike.giles@maths.ox.ac.uk}{0000-0002-5445-3721}{University of Oxford}{Mathematical Institute}{Andrew Wiles Building, Radcliffe Observatory Quarter Woodstock Road}{Oxford}{OX2 6GG}{UK}
\paperauthor{Wesley~Armour}{wes.armour@oerc.ox.ac.uk}{0000-0003-1756-3064}{University of Oxford}{Oxford e-Research Centre}{Oxford}{}{OX1 3QG}{UK}

\begin{abstract}
We present an improvement of our implementation of the Correlation Technique for the Fourier Domain Acceleration Search (FDAS) algorithm on Graphics Processor Units (GPUs) \citep{2015arXiv151107343D,Dimoudiinprep}. Our new improved convolution code which uses our custom GPU FFT code is between 2.5 and 3.9 times faster the than our cuFFT-based implementation (on an NVIDIA P100) and allows for a wider range of filter sizes then our previous version. By using this new version of our convolution code in FDAS we have achieved 44\% performance increase over our previous best implementation. It is also approximately 8 times faster than the existing PRESTO GPU implementation of FDAS \citep{PrestoGPU}. This work is part of the AstroAccelerate project \citep{AstroAccelerateGit}, a many-core accelerated time-domain signal processing library for radio astronomy.
\end{abstract}

\section{Introduction}
Binary pulsars are an important target for radio surveys because they present a natural laboratory for testing general relativity. The orbital motion of pulsars locked in a binary system causes a frequency shift (a Doppler shift) in their normally strictly periodic pulse emissions. These shifts cause a reduction in the sensitivity of traditional periodicity searches to a point where some accelerated pulsars cannot be detected. To partially correct for this effect and hence allow observation of these systems a set of signal processing methods called acceleration searches are used. There are two routinely used methods. First is TDAS (time domain acceleration search), which is based on re-sampling data at a given trial acceleration value followed by Fourier transforming the data using an FFT. The second method is called the Fourier domain acceleration search or FDAS. This work is a continuation of our GPU implementation of FDAS \citet{Dimoudiinprep}. By referring to our old code we refer to the code presented in \citet{Dimoudiinprep}.

The acceleration searches assume a constant acceleration $a$ over a duration of the observation. The constant acceleration $a$ corresponds to a constant frequency derivative $\dot{f}$ which is related to the number of Fourier bins $z$, that the signal frequency has drifted during the observation time $T$:
\begin{equation}
a=\frac{\dot{f}}{f_0}c=\frac{zc}{f_{0}T^2}\,.
\end{equation}
The Fourier domain acceleration search (FDAS) is based on the correlation technique \citep{2002AJ....124.1788R}, which works by correlating a predicted Fourier response that corresponds to given frequency derivative or $z$ number with the FFT output of the observation. This is effectively a matched filtering process. The corrected Fourier response $r_0$ of the signal at frequency bin $r_0$ is recovered by correlating the $m$ bins around bin $r_0$ with the predicted, normalized, Fourier response $H$ to a given orbital acceleration as follows:
\begin{equation}
\label{matchedfiltering}
A_{r0}=\sum^{r0+m/2}_{k=r0-*m/2}A_{k}H^{*}_{r0-k}\,.
\end{equation}
Given the computational complexity of the FDAS algorithm and its importance in discovering pulsar binaries any performance improvement is valuable.

This implementation of the FDAS algorithm is part of AstroAccelerate library \citet{AstroAccelerateGit}. The AstroAccelerate library is a GPU software package that focuses on enabling real-time processing of time-domain radio astronomy data. It uses the CUDA programming language for NVIDIA GPUs. The massive computational power of modern day GPUs allows the code to perform algorithms such as de-dispersion \citet{2012ASPC..461...33A} or single pulse searching \citet{2016arXiv161109704A} in real-time on very large data-sets which are comparable to those which will be produced by next generation radio-telescopes such as the SKA.

\section{Improvements}
	The most computationally intensive part of FDAS is a convolution routine which takes care of the matched filtering process \eqref{matchedfiltering}. The convolution itself is done via the overlap-and-save method which divides the long input time series into smaller segments and performs Fourier domain convolution on each of those segments independently. These segments are then merged in such a way as to produce the linear convolution. More information can be found in \citet{Dimoudiinprep}. The convolution code itself can be separated into two components: the custom FFT code which performs the FFT on data resident in a GPUs shared memory and then the  convolution in the Fourier domain. This work improves the performance of both. In this section we describe improvement in performance which do not affect generality of the our convolution code. Our final result also includes FDAS specific optimizations, which cannot be used for general convolution as they exploit symmetries in FDAS algorithm (more can be found in \citet{Dimoudiinprep}).
	
First we will describe the improvements in our custom FFT code. The discrete Fourier transform is given by the formula:
	\begin{equation}
	\label{eqa:DFT}
		X_m=\sum_{n=0}^{N-1} x_n \textup{e}^{-i2\pi nm/N}=\sum_{n=0}^{N-1} x_n\, \omega_{N}^{nm}\,,
	\end{equation}
	where $X_m$ is a signal in the \textit{frequency-domain}, $x_n$ is a signal in the \textit{time-domain},  $N$ is the signal length or FFT length and the exponential factors $\omega_{N}^{nm} = \textup{e}^{-\textup{i}2\pi nm/N}$ are called \textit{twiddle} factors. The algorithm behind the FFT is based on a divide and conquer paradigm, each FFT is itself composed from smaller FFTs which are in turn composed from even smaller FFTs. These FFTs are independent of one another until they are merged to form a bigger FFT. This gives the opportunity for significant data reuse as each of these FFT uses the same twiddle factors. Thus by calculating more than one FFT element per GPU thread and by arranging that each of these elements form an independent FFT we can reuse already calculated twiddle factors. However increasing the number FFT elements per thread consumes more and more GPU resources and eventually leads to a decrease in performance.	
	
	We also take advantage of the GPU shuffle instructions. These instructions allow a warp (a set of 32 threads which are always synchronized) to share data very quickly through registers. To make use of shuffles we have divided the FFT calculation into two steps. The first step calculates the FFT up to size $N=32$ using shuffle instructions, the advantage of this is not only the quick exchange of data but also eliminates the need to synchronize threads. The second step performs the remainder of the FFT in the range $N>32$. Our new custom FFT implementation halves the number of shuffle instructions used, this together with increasing the number of FFT elements calculated per thread in both parts of the code (to four) has given our reported improvement in performance of our custom FFT code. These improvements have increased performance of our custom FFT by 20\% in case of P100 GPU and 50\% in case of TITAN X GPU. Since the FFT algorithm is called multiple times per segment any improvement has a major impact on the run time of the whole convolution code.

\section{Results}

In a convolution code that uses the overlap and save method our custom FFT will be called from within a threadblock (i.e. it executes on data that is already resident in shared memory) and it will be called for every matched filter that the input signal is to be convolved with. This makes our convolution code (which uses our custom FFT code) shared memory bandwidth bound. This is different to a convolution code that uses cuFFT because cuFFT would need to write back to global memory for every matched filter that the input data is convolved with, making it global memory bandwidth bound. To this end the results presented below come from calculating 100 FFTs per kernel launch, while launching 1000 of these kernels thus in total we performing 100 000 FFT of the given length. The results of our comparison for an FFT of length $N=1024$ are presented in table \ref{tab:FFTtimes}. It is important to note that the fastest FFT code might not produce the fastest convolution code because using our FFT code within an existing kernel will change parameters such as register usage and occupancy of the given kernel.

\begin{table}[!ht]
\caption{Comparison of our custom FFT kernels for $N=1024$. The number of elements processed by each thread per one iteration is in brackets. The slowdown of the shuffle code is due to the large number of twiddle factors computed by the kernel.}
\label{tab:FFTtimes}
\smallskip
\begin{center}{
\small
\begin{tabular}{lcc}
\tableline
\noalign{\smallskip}
Code variant & P100 & TITAN X (p) \\
& [ms] & [ms] \\
\noalign{\smallskip}
\tableline
\noalign{\smallskip}
	FFT Sm. Mem. (2 elem.)  & 2.14 & 2.39 \\
	FFT Sm. Mem. (4 elem.) & 1.80 & 2.25 \\
	FFT Sm. Mem. (8 elem.) & 2.12 & 2.21 \\
	FFT Sm. Mem. old & 2.17 & 3.50 \\
	cuFFT (Gl. mem. limited) & 3.05 & 4.70 \\
\noalign{\smallskip}
\tableline
\end{tabular}
}
\end{center}
\end{table}

Comparison of our new improved custom FFT convolution code to our cuFFT convolution code is summarized in table \ref{tab:CONVtimes}. Our new custom FFT code also allowed us to extend the segment size up to 2048 elements which means we can now use longer templates for matched filtering.

\begin{table}[!ht]
\caption{Comparison of averaged speedups of our new custom FFT convolution code to our older codes for different GPUs. Results without brackets give the speedup compared to our cuFFT convolution, results in square brackets give the speedup compared to our old custom FFT convolution code.}
\label{tab:CONVtimes} 
\smallskip
\begin{center}{
\small
\begin{tabular}{ccc}
\tableline
\noalign{\smallskip}
Filter length & P100 & TITAN X (p) \\
\noalign{\smallskip}
\tableline
\noalign{\smallskip}
	64  & 3.89 [1.28] & 4.48 [1.30] \\
	128 & 3.45 [1.30] & 4.03 [1.35] \\
	256 & 2.94 [1.37] & 3.57 [1.35] \\
	384 & 2.60 [1.37] & 3.36 [1.39] \\
	512 & 2.45 [1.36] & 3.16 [1.37] \\
\noalign{\smallskip}
\tableline
\end{tabular}
}
\end{center}
\end{table}

When we use our new convolution code in our FDAS code we see performance increase by 44\%. The FDAS code performs better than the presented results of our general convolution code since the new convolution code allows for further optimizations which are specific for FDAS algorithm (for example it allows us to reduce GPU register usage).

\section{Conclusions}
Our continuous improvements have increased performance of our FDAS code by 44\%. This was achieved by optimizing our custom FFT code as well as improving our convolution code. The improved performance of our FDAS code means that we can process more data per GPU thus offering cost savings because fewer GPUs are needed to perform the same task (and also less energy is used) or we are able to search the parameter space in finer detail allowing for more detections and improvements in the signal-to-noise of outputted results.


\begin{thebibliography}{}
\expandafter\ifx\csname natexlab\endcsname\relax\def\natexlab#1{#1}\fi
\expandafter\ifx\csname url\endcsname\relax
  \def\url#1{\texttt{#1}}\fi
\expandafter\ifx\csname urlprefix\endcsname\relax\def\urlprefix{URL }\fi
\providecommand{\eprint}[2][]{\url{#2}}

\bibitem[{{Ad\'{a}mek} \& {Armour}(2016)}]{2016arXiv161109704A}
{Ad\'{a}mek}, K., \& {Armour}, W. 2016, in ADASS XXVI, edited by F.~{Pasian}.
  \eprint{1611.09704}

\bibitem[{{Armour} et~al.(2002)}]{AstroAccelerateGit}
{Armour}, W., et~al. 2002, Astroaccelerate,
  \url{https://github.com/AstroAccelerateOrg}

\bibitem[{{Armour} et~al.(2012)}]{2012ASPC..461...33A}
--- 2012, in ADASS XXI, edited by P.~{Ballester}, D.~{Egret}, \& N.~P.~F.
  {Lorente}, vol. 461 of ASP Conf. Ser., 33. \eprint{1111.6399}

\bibitem[{{Dimoudi} \& {Armour}(2015)}]{2015arXiv151107343D}
{Dimoudi}, S., \& {Armour}, W. 2015. \eprint{1511.07343}

\bibitem[{{Dimoudi} et~al.(2017)}]{Dimoudiinprep}
{Dimoudi}, S., et~al. 2017, submitted to ApJS

\bibitem[{{Luo}(2013)}]{PrestoGPU}
{Luo}, J. 2013, Presto gpu, \url{https://github.com/jintaoluo/presto_on_gpu}

\bibitem[{{Ransom} et~al.(2002){Ransom}, {Eikenberry}, \&
  {Middleditch}}]{2002AJ....124.1788R}
{Ransom}, S.~M., {Eikenberry}, S.~S., \& {Middleditch}, J. 2002, The
  Astronomical Journal, 124, 1788. \eprint{astro-ph/0204349}

\end{thebibliography}

\end{document}